\documentclass[a4paper]{article}
\usepackage[utf8x]{inputenc} 
\usepackage{graphicx}
\usepackage{hyperref}
\usepackage{color}
\usepackage[margin=1.5in]{geometry}
\usepackage[table,xcdraw]{xcolor}
\usepackage[labelfont=bf]{caption}
\usepackage{caption}
\usepackage{float}
\usepackage{authblk}
\usepackage{datetime}
\usepackage{adjustbox}
\usepackage{amssymb}
\usepackage{amsmath}
\title{Deep Reinforcement Learning for Asset Allocation in US Equities}
\author[1]{Dr Miquel Noguer i Alonso}
\author[2]{Sonam Srivastava}

\affil[1]{Artificial Intelligence in Finance Institute, NYU Courant}
\affil[2]{Wright Research}

\date{October 2020}

\begin{document}

\maketitle

\begin{abstract}

Reinforcement learning is a machine learning approach concerned with solving dynamic optimization problems in an almost model-free way by maximizing a reward function in state and action spaces. This property makes it an exciting area of research for financial problems. Asset allocation, where the goal is to obtain the weights of the assets that maximize the rewards in a given state of the market considering risk and transaction costs, is a problem easily framed using a reinforcement learning framework. So it is first a prediction problem for the vector of expected returns and covariance matrix and then an optimization problem for returns, risk, and market impact, usually a quadratic programming one.
Investors and financial researchers have been working with approaches like mean-variance optimization, minimum variance, risk parity, and equally weighted and several methods to make expected returns and covariance matrices' predictions more robust and after use mean-variance like the Black Litterman model. 
This paper demonstrates the application of reinforcement learning to create a financial model-free solution to the asset allocation problem, learning to solve the problem using time series and deep neural networks. 
We demonstrate this on daily data for the top 24 stocks in the US equities universe with daily rebalancing. We use a deep reinforcement model on US stocks using different deep learning architectures. We use Long Short Term Memory networks, Convolutional Neural Networks, and Recurrent Neural Networks and compare them with more traditional portfolio management approaches like mean-variance, minimum variance, risk parity, and equally weighted. The Deep Reinforcement Learning approach shows better results than traditional approaches using a simple reward function and only being given the time series of stocks.
In Finance, no training to test error generalization results come guaranteed. We can say that the modeling framework can deal with time series prediction and asset allocation, including transaction costs. 

\end{abstract}
\section{Introduction}

Asset allocation is one of the most critical problems in Finance, computing the optimal weights of assets in a portfolio for a given investment horizon. The expected returns and risks for this period are the ingredients needed for traditional asset allocation models. We will investigate in this paper if we can construct portfolios using a reinforcement learning framework.

\section{Asset Allocation Methodologies}
Asset allocation or portfolio selection is the process of finding optimal weights for component assets in a portfolio is one of the most significant areas of research in modern Finance. Traditional asset allocation implementations involve several modeling steps: First, estimating a vector of expected return for the asset universe, second estimating the covariance matrix, and then optimizing the portfolio allocation minimizing the market impact using one of the methodologies described in this section. So it is first a prediction problem for the vector of expected returns and covariance matrix and then an optimization problem, usually a quadratic programming one.

This economic research field has evolved over the past 50 years, from the original 60/40 portfolio, modern portfolio theory in the 50's to other approaches that partially fix the MPT flaws.

Modern portfolio theory by Markovitz \cite{markowitz} mean-variance optimization maximizes the expected return of a portfolio. Risk and return trade-off is at the heart of the MVO theory, designed to optimize a portfolio for a single period. The ingredients needed are the vector of expected returns and the covariance matrix.
Mean-variance optimizations with affine constraints and linear market impact is a quadratic programming problem, and we can solve it semi-analytically.
A few years after \cite{10.2307/2328079} show that if the returns on all assets available for portfolio formation are jointly elliptically distributed, then all portfolios can be characterized entirely by their location and scale. Any two portfolios with exact location and scale of portfolio return have identical distributions of portfolio return. If distributions are not elliptical, then mean - conditional value at risk \cite{Rockafellar00optimizationof} the right thing to do.

MPT has several flaws: the non-elliptical case solved using mean conditional value at risk, sensitivity to model inputs ( expected returns and covariance matrix), and model inputs are indeed very hard to estimate well. Researchers and practitioners have devised several techniques to estimate expected returns and the covariance matrix. Black and Litterman developed the Black Litterman model \cite{blacklitterman}. The model uses equilibrium assumptions. The user states how his assumptions about expected returns differ from the markets and state his degree of confidence in the alternative assumptions. From this, the Black–Litterman method computes the (mean-variance efficient) asset allocation.

In addition to BL, other methodologies use risk like the equally weighted portfolio, the 1/n portfolio, the equal volatility portfolio that uses the same amount of volatility in every asset, the Minimum Variance Portfolio, and the ubiquitous Risk Parity. Portfolio Risk Parity \cite{riskparity} (using the same amount of marginal contribution to risk) and  Maximum Sharpe Ratio Portfolio. The methodologies using risk implicitly assume that the Sharpe ratio will be equal in all assets.

Estimation can be improved using several statistical methods. For the vector of expected returns, the starting point is the historical returns, Exponentially weighted mean historical returns, the James-Stein Shrinkage \cite{james1961} as well as equilibrium returns a-la Black-Litterman. The covariance matrix's starting point is the sample covariance matrix and some covariance shrinkage, Non-linear shrinkage estimation of large-dimensional covariance matrices \cite{ledoit2012} and the minimum covariance estimator \cite{article}. Meucci \cite{meucci2010fully} has also made exciting contributions like the entropy pooling model.

With the increasing use of artificial intelligence methods in Finance, machine learning, or deep learning models have been used extensively in a supervised or unsupervised fashion for this problem statement, mostly for expected return prediction, some other researchers have worked on the covariance matrix. 

The emergence of big data and powerful computers gave rise to machine learning algorithms that have shown significant success in image detection and recognition. Further success was shown in natural language processing using deep neural networks. Reinforcement learning algorithms are, in turn, proving to be a powerful tool in several applications across numerous fields, ranging from traffic light control to robotics, optimizing chemical reactions, advertising, and gaming. 

Several authors have made important contributions to the use and research of machine learning algorithms in finance like\cite{de2020machine}, \cite{miquel2018}, \cite{book} and \cite{coqueret2020machine} to cite a few good references.
Given the complex nature of the financial world, financial modelers should consider performance in terms of the generalization error. However, they should also consider other essential aspects, such as interpretability, fairness, ethics, and privacy. 

Researchers have showcased increased accuracy in prediction using sophisticated deep neural networks like Recurrent Neural Networks and Long Short Term Memory models in a supervised setup \cite{miquel2018}. Nevertheless, the problem is a challenging one as the predictions are weak due to the problematic nature of predicting future market prices.  After a supervised learning model, we need to compute the optimal weights following one of the methodologies described above.
We introduce a pure machine learning-driven framework for the portfolio management problem that uses reinforcement learning for asset allocation.

As \cite{ritter2017machine} describe in their seminal research, problems like portfolio management are dynamic optimization problems to determine the
best actions possible that maximize the relative value between two or more
payoffs at different points in time. The most common approach for
solving dynamic optimization problems of this kind is dynamic programming (DP). However, due to the enormous computational scale of the problem, DP is infeasible. Reinforcement learning (RL) that follows a simple "trial and error" by receiving feedback using the amount of reward resulting from each action it takes is a much more feasible way of solving such problems.

The reinforcement learning approach also has an added advantage of learning based on a custom domain-specific reward function as opposed to \textit{accuracy} of prediction or similar rewards used by supervised learning methods that ignore the problem. An example is controlling the portfolio turnover or excessive trading due to noisy signals that drive down the portfolio returns. 

\section{Reinforcement Learning}

Optimal control of unknown Markov decision processes is how one defines the problem of reinforcement learning. A learning agent tries to capture the most important aspects of a real problem interacting with its environment over time. The agent senses the state of its environment and takes actions that affect the state. The agent tries to maximize a goal relating to the state of the environment. \cite{Sutton1998}

Speaking Mathematically, RL is a way to solve multi-period optimal control problems. The RL agent's policy typically consists of explicitly maximizing the action-value function for the current state. This value function is an approximation of the actual value function of the multi-period optimal control problem. Training refers to improving upon the approximation of the value functions as more training examples are made available.

Reinforcement learning is different from supervised learning in which the agent learns based on correctly labeled datasets so that it can predict correctly in examples out of the training set. Supervised learning solutions are typically not suitable for an interactive problem where the agent learns from experience. Reinforcement learning differs from unsupervised learning, typically about finding structure or representation of unlabeled data without an explicit reward or question in mind.

A key feature of reinforcement learning is that it models an agent's whole task of maximizing rewards interacting with an uncertain environment. This feature contrasts with others that consider subtasks generally addressing how they might fit into a larger picture sequentially.

Unlike other machine learning frameworks, the reinforcement learning agent also balances the trade-off between exploration and exploitation while solving a problem. This approach is goal-seeking towards an explicit reward, which is a step in using simple principles in machine learning.

Reinforcement learning allows us to solve these dynamic optimization tasks in a close to "model-free" way, relaxing the assumptions often needed for dynamic programming (DP) approaches. As mentioned previously, modeling stochastic dynamics is an open and challenging problem. In addition to that, modeling market impact and transaction costs is not a simple task due to their non-linear and non-differentiable behavior. 

Reinforcement Learning (RL)  is concerned with using data and experiments and learning an acceptable policy or strategy for the agent with relatively simple feedback. With the optimal strategy, the agent can adapt to the environment to maximize future rewards actively.
The agent is acting in an environment. How the environment reacts to specific actions is defined by a model that initially might be known or not. The agent can stay in a given state ($ s \in \mathcal{S}$) of the environment. It can choose to take action ($a \in \mathcal{A}$) to switch from one state to another. The transition probabilities between states (P) govern the state the agent arrives at. Once an action is taken, the environment might deliver an instant or a delayed reward ($r \in \mathcal{R}$).

The reward function and transition probabilities define the model.  There are two situations based on how much we know about the model:
\begin{itemize}
\item Know the model: planning with perfect information; do model-based RL. 

\item Model unknown: earning with incomplete information; do model-free RL or try to learn the model explicitly as part of the algorithm. 
\end{itemize}

The agent's policy $\pi(s)$ guides the optimal action to take in a given state to maximize total rewards. There is a value function $V(s)$ predicting the expected amount of future rewards associated with each state that we can receive in this state by following a given policy. Both policy and value functions are what reinforcement learning tries to learn.

The interaction between the environment and the agent involves a sequence of actions and observed rewards in time, $t=1, 2, \dots, T$. During the process, the agent accumulates knowledge about the environment, learns the optimal policy, and makes decisions on which action to take next to learn the best policy efficiently. Let’s call the state, action, and reward at time step t as $S_t$, $A_t$, $R_t$ , respectively. The interaction sequence is fully described by an episode and the sequence finishes at the terminal state $S_T$ : $S_1$, $A_1$, $R_1$ ,  $S_2$, $A_2$,$\dots$, $S_T$
 
Terms one will encounter a lot when diving into different categories of RL algorithms:

\begin{itemize}

\item Model-based: Rely on the environment; either the model is known, or the algorithm learns it explicitly.
\item Model-free: No dependency on the model during learning.
\item On-policy: Use the deterministic outcomes or samples from the target policy to train the algorithm.
\item Off-policy: Training on the distribution of transitions or episodes produced by a different behavior policy rather than produced by the target policy.
\end{itemize}

\paragraph{Model transition and reward}

The model tries to describe the environment. With the model, we can infer how the environment would interact with and provide feedback to the agent. The model has two significant parts, transition probability function $P$ and reward function $R$.

Let us say when we are in state $s$, and we decide to take action to arrive in the next state, $s\text{'}$ and obtain reward $r$. This is known as a transition step, represented by a tuple ($s, a, s\text{'}, r$).

The transition function P is the probability of transitioning from state $s$ to $s\text{'}$ after taking action $a$ and obtaining reward $r$. We use $P$ as a symbol of "probability."

\begin{align}
P(s', r \vert s, a)  = \mathbb{P} [S_{t+1} = s', R_{t+1} = r \vert S_t = s, A_t = a]
\end{align}

The state-transition function is defined as  \begin{align}
P(s′,r|s,a)
\end{align}

\begin{align}
P_{ss'}^a = P(s' \vert s, a)  = \mathbb{P} [S_{t+1} = s' \vert S_t = s, A_t = a] = \sum_{r \in \mathcal{R}} P(s', r \vert s, a)
\end{align}

The reward function R predicts the reward triggered by the action:
\begin{align}
R(s, a) = \mathbb{E} [R_{t+1} \vert S_t = s, A_t = a] = \sum_{r\in\mathcal{R}} r \sum_{s' \in \mathcal{S}} P(s', r \vert s, a)
\end{align}

\paragraph{Policy}

Policy, as the agent’s sequence of actions function $\pi$, tells us which action to take in state s. It is a function from state $s$ to action $a$ and can be stochastic or deterministic:

\begin{itemize}
\item Stochastic: $\pi(a|s)=P\pi[A=a|S=s]$
\item Deterministic: $ \pi(s)=a $.
\end{itemize}

\paragraph{Value Function}

Value function measures the future reward of a state or an action. The return or future reward is the total sum of discounted rewards. Let’s compute the return $Gt$ The discounting factor $\gamma$ ∈[0,1] takes care of future rewards and its uncertainty:
\begin{align}
G_t = R_{t+1} + \gamma R_{t+2} + \dots = \sum_{k=0}^{\infty} \gamma^k R_{t+k+1}
\end{align}
We can define the action-value or Q-value of a state-action pair as:

\begin{align}
Q_{\pi}(s, a) = \mathbb{E}_{\pi}[G_t \vert S_t = s, A_t = a]
\end{align}
As we are following the target policy $\pi$, we can make use of the probability distribution over Q-values and possible actions to recover the state-value:

\begin{align}
V_{\pi}(s) = \sum_{a \in \mathcal{A}} Q_{\pi}(s, a) \pi(a \vert s)
\end{align}
The action advantage function ( A-value) is the difference between action-value and state-value:
\begin{align}
A_{\pi}(s, a) = Q_{\pi}(s, a) - V_{\pi}(s)
\end{align}

\paragraph{Optimal Value and Policy}

The optimal value function achieves the maximum return:
\begin{align}
V_{*}(s) = \max_{\pi} V_{\pi}(s),
Q_{*}(s, a) = \max_{\pi} Q_{\pi}(s, a)
\end{align}

The optimal policy offers optimal value functions:

\begin{align}
\pi_{*} = \arg\max_{\pi} V_{\pi}(s),
\pi_{*} = \arg\max_{\pi} Q_{\pi}(s, a)
\end{align}
And we have $V_{\pi_{*}}(s)=V_{*}(s)$ and $Q_{\pi_{*}}(s, a) = Q_{*}(s, a)$

\paragraph{Belman Equations}

Bellman equations are the set of equations that decompose the value function into the immediate reward and future discounted values.

\begin{align}
V(s) &= \mathbb{E}[G_t \vert S_t = s] \\
&= \mathbb{E} [R_{t+1} + \gamma R_{t+2} + \gamma^2 R_{t+3} + \dots \vert S_t = s] \\
&= \mathbb{E} [R_{t+1} + \gamma (R_{t+2} + \gamma R_{t+3} + \dots) \vert S_t = s] \\
&= \mathbb{E} [R_{t+1} + \gamma G_{t+1} \vert S_t = s] \\
&= \mathbb{E} [R_{t+1} + \gamma V(S_{t+1}) \vert S_t = s] 
\end{align}
For Q-Value

\begin{align}
Q(s, a) 
&= \mathbb{E} [R_{t+1} + \gamma V(S_{t+1}) \mid S_t = s, A_t = a] \\
&= \mathbb{E} [R_{t+1} + \gamma \mathbb{E}_{a\sim\pi} Q(S_{t+1}, a) \mid S_t = s, A_t = a]
\end{align}

\paragraph{Solving RL problems}

When we know the model follows Bellman equations, we can use Dynamic Programming (DP) to evaluate value functions iteratively and improve policy. We can use Monte-Carlo methods,  Temporal-Difference (TD) Learning,  Q-learning, and policy gradient to follow a model-free approach. The Actor-Critic algorithm gets the value function in addition to the policy.

\begin{itemize}
\item Critic: updates value function parameters $w$ and depending on the algorithm it could be action-value $Q(a|s;w)$ or state-value $V(s;w)$.
\item Actor: updates policy parameters $\theta$, in the direction suggested by the critic, $\pi(a|s;\theta)$.
\end{itemize}

\section{Reinforcement Learning in Finance Literature Review}

Several authors have explored the use of reinforcement learning techniques in Finance in different areas like option delta hedging,  trading,  tax optimization for some areas like portfolio allocation. The current literature on Reinforcement Learning in trading is using technically mainly these three methods: critic-only, actor-only, and actor-critic approach. 

The critic-approach, mainly Deep Q Network, is the most published method in this field where a state-action value function, Q, is constructed to represent how good a particular action is in a state. 

The second most common approach is called the actor-only, where the agent directly optimizes the objective function without computing the expected rewards of each action in a state. Actor-only approaches are generalized to continuous action spaces as they learn the policy directly. The Actor-only approach is different from standard RL approaches, where the policy needs the distribution to be learned. The Policy Gradient Theorem and Monte-Carlo methods are used to study the distribution of a policy in training, and models are updated until the end of each episode.
These models often experience slow learning and need many samples to obtain an optimal policy as individual wrong actions will be considered acceptable as long as the total rewards are good, taking a long time to adjust these actions. We will explore solutions to that in future research.

The third method is the actor-critic, which models learning problems by updating the policy in real-time. The key idea is to update two models - the actor and the critic. The actor decides how an agent performs in the current state, and the critic measures how good the action is.

While all these methods are useful in different contexts in finance portfolio allocation, the reward function and value function can be as simple as the next period returns.

The main contributions to the area in our view are:
\begin{itemize}

\item \cite{ritterkolm} gives an overview of reinforcement learning in financial applications. In Finance, the intertemporal choice includes pricing and hedging options, trading, portfolio allocation subject to transaction costs, market making, asset-liability management, client management, and tax optimization. Reinforcement learning allows agents to solve these dynamic optimization problems in an almost model-free way, changing the assumptions used in classical financial models more flexibly.
\item Following \cite{ritterkolm} approach, some of the most critical problems in Finance can be cast as a value function. Option hedging, optimal execution, and optimal trading of alpha forecasts all share the property's value function is the expected integrated revenue and variance.

\begin{align}
    v_{\pi}(s)= \mathbb{E} \bigg [ \int_0^T \big(x_t \, r_t - \frac{\lambda}{2} \, x_t^2 \sigma_t^2 - f(\dot{x})\big) \mathrm{d} t \, \bigg  | \, x_0 = s \bigg],
\end{align}

where $f(\dot{x})$ is some function of the time-derivative $\dot{x}_t = \frac{\mathrm{d}x_t}{\mathrm{d} t}$ approximating market impact.
\item \cite{book} the authors introduce a probabilistic extension of Q-learning known in the literature as "G-learning." As known previously, when a reward function is quadratic, even including market impact, neither approach is needed as portfolio optimization procedure is semi-analytic. If we consider the non-linear market impact, the reward function turns into non-quadratic, then numerical methods are needed. They also cover Inverse Reinforcement learning (IRL) and imitation learning (IL). These methods solve the problem of optimal control in a data-driven way, similarly to reinforcement learning. The difference is that modelers do not know the reward function. The problem is learning the reward function from the observed behavior of an agent.

\item \cite{moody} present an adaptive algorithm called recurrent reinforcement learning (RRL) for discovering investment policies. They demonstrate how direct reinforcement can optimize risk-adjusted investment returns (including the differential Sharpe ratio) while accounting for transaction costs.

\item \cite{dempster} introduces adaptive reinforcement learning (ARL) as the basis for a fully automated trading system application. The system is designed to trade foreign exchange (FX) markets. It relies on a layered structure consisting of a machine learning algorithm, a risk management overlay, and a dynamic utility optimization layer.

\item \cite{deng} proposes a task-aware backpropagation through time method to cope with the gradient vanishing issue in in-depth training. The neural system's robustness is verified on both the stock and the commodity futures markets under general testing conditions.

\item \cite{hens}  apply the recurrent reinforcement learning method of \cite{moody} in the context of the strategic asset allocation computed for sample data from the United States, the United Kingdom, and Germany. They show that the investor actively times the market, and he can outperform it consistently concerning risk-adjusted returns over the almost two decades analyzed.

\item \cite{Zhang2019DeepRL} The paper explores the use of reinforcement learning algorithms to trade 50 liquid contracts using Deep Q Learning, Policy Gradients, and advantage Actor-Critic in discrete and continuous action spaces using time-series momentum and technical indicators as action spaces.

\item An excellent reinforcement learning implementation in portfolio management can be found in \cite{jiang2017deep}. The authors use the framework to optimize a portfolio's performance consisting of crypto-currencies and show exciting results. These researchers also introduced portfolio vector memory for controlling turnover that we also use in this work.

\item An excellent reference to understand the properties of deep learning from a mathematical standpoint \cite{e2020mathematical} provide an excellent review of the current mathematical understanding of deep neural networks. 
\end{itemize}

\section{Asset Allocation using Reinforcement Learning}

Portfolio management can be modeled in the reinforcement-learning framework as follows - the \textit{agent} is the neural network of choice along with the portfolio memory component, the \textit{state} it acts upon is the market snapshot at the given point in time as represented by a tensor of features relating to the component assets. The \textit{action} is the final portfolio weights that the framework gives out for the given period. The \textit{environment} that receives this action is the market, which sends out the cost adjusted returns for the period as \textit{reward} back to the neural network agent. The agent uses gradient descent based on the reward to learn.
When agents operate in environments on which they have limited knowledge, we need a new modeling paradigm to overcome this. Reinforcement-learning offers us a range of tools to deal with agents that have limited knowledge of the environment.

\section{Our methodology}

As described in this paper, the reinforcement learning agent tries to maximize the portfolio vector memory's cost adjusted returns using a network component called the portfolio vector memory that stores old portfolio weights and motivates the network to reduce the churn to reduce costs. This concept was introduced in \cite{jiang2017deep}, where they try to find the best method for trading cryptocurrencies based on the reinforcement learning approach.
The model's architecture consists of four components - an input tensor containing a history input features, a neural network that predicts an expected return vector, a portfolio vector memory that incorporates the historical weights in the portfolio, and a final softmax convolution that predicts the final weights. The final weights go to the market (or the environment in RL nomenclature) and receive the rewards in terms of cost-adjusted returns fed back to the network that uses gradient descent to learn based on the reward. We use a convolutional neural network, recurrent neural network, and a long short term memory network as our predictor networks and compare performance.

We demonstrate that this approach compares well to traditional and popular portfolio construction methods like an equal-weighted portfolio, Markovitz mean-variance optimal portfolio \cite{markowitz}, and the Risk Parity Portfolio \cite{riskparity}.

\subsection{Data}
We use the daily bar OHLC data of top 24 US equity securities by market cap. The data is available for the two years between \newdate{date}{01}{01}{2008} \date{\displaydate{date}} to \newdate{date1}{1}{06}{2020} \date{\displaydate{date1}}. We trade with a investment horizon of 1 day. 75\% of the data is taken to train the network and 25\% for testing. More details on the rolling training and testing framework in \ref{topic:agent}.

The stocks used are:
\begin{table}[h]
\centering
\begin{tabular}{lllll}
\rowcolor[HTML]{EFEFEF} 
AAPL & AMZN & BAC & BRK A & CVX \\
DIS & FB & GOOG & HD & INTC \\
\rowcolor[HTML]{EFEFEF} 
JNJ & JPM & KO & MA & MRK \\
MSFT & PFE & PG & T & UNH \\
\rowcolor[HTML]{EFEFEF} 
V & VZ & WFC & WMT & XOM \\
\end{tabular}
\caption{List of stocks used for the research }
\label{table:stocks}
\end{table}

\subsection{Reinforcement Learning Framework}

Reinforcement Learning can be represented as a simple Markov Decision Process where an \textit{agent} sees a given \textit{state} based on which it takes an \textit{action} which goes to the \textit{environment} that sends back a \textit{reward} to the agent based on which it learns.

\begin{figure}[H]
    \centering
    \captionsetup{justification=centering}
    \includegraphics[width=300pt]{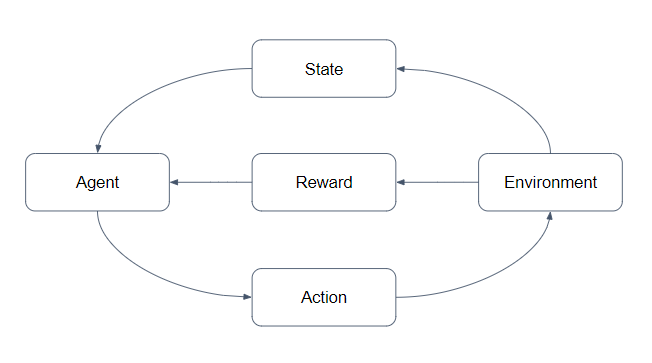}
    \caption{Basic representation of the reinforcement learning framework as a Markov Decision Process.}
    \label{fig:rl_framework}
\end{figure}

The problem of portfolio management can be modeled into the reinforcement learning framework. The \textit{agent} is the neural network of choice along with the portfolio memory component. The \textit{state} it acts upon is the market snapshot at the given point in time, as represented by a tensor of features relating to the component assets. The \textit{action} is the final portfolio weights that the framework gives out for the given period. The \textit{environment} that receives this action is the market, which sends out the cost adjusted returns for the period as \textit{reward} back to the neural network agent. The agent uses gradient descent based on the reward to learn.

\begin{figure}[H]
    \centering
    \captionsetup{justification=centering}
    \includegraphics[width=300pt]{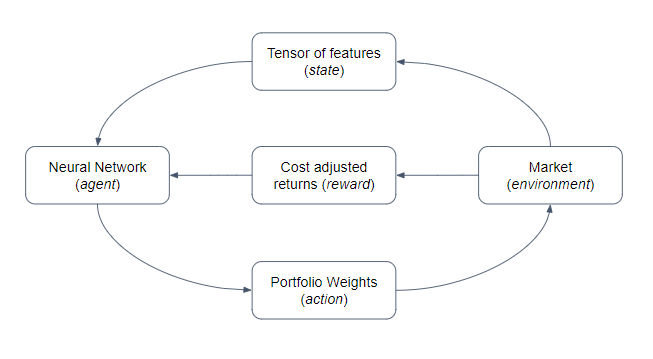}
    \caption{Portfolio management problem represented as a reinforcement learning framework.}
    \label{fig:pm_framework}
\end{figure}

\subsection{State - Input Tensor}

We take high, low, and close (HLC) for 50 periods for the 24 stocks to create the (3 x 50 x 24) input tensor. More features, if available, can be added as input here.

\begin{figure}[H]
    \centering
    \captionsetup{justification=centering}
    \includegraphics[width=250pt]{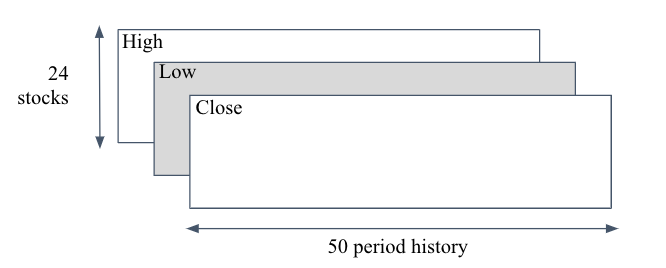}
    \caption{The input to the network is the price tensor for 50 periods consiting of high low and close prices for the 24 stocks}
    \label{fig:input_framework}
\end{figure}

\subsection{Agent - Neural Network} \label{topic:agent}

As represented by the input tensor, the agent acting based on the state is the neural network. We consider three neural network variations - a convolutional neural network, a recurrent neural network, and a long short-term memory neural network.

\paragraph{Sequential Mini Batch Training}
Using the full data for training at each time step would make the computations time insensitive and complex. We train the network based on sequential mini-batches. Unlike the usual way of picking random mini-batches, we pick mini-batches for training in their time-order. So at any given time period $t$ the network is trained on $HLC$ prices for the 24 stocks from $t - n - 1$ to $t - 1$ where $n$ is the mini-batch size, 50 in our case.

The three network topologies we have experimented with are introduced below.

\subsubsection{Convolutional Neural Network}
Convolutional neural networks are variants of neural networks inspired by neurons in the visual cortex that enable vision, where each neuron only processes data for its receptive field \cite{refCNN}. These networks consist of a set of convolutional layers that \textit{convolve} the input using a dot product followed by a ReLU activation function and a max-pooling layer or fully connected layers. One final convolution pushes out the output from the network. At each convolution, the network extracts a feature, like a neuron recognizing an edge or a shape in vision. The pooling layers streamline the cluster of outputs from multiple convolutions to single output using operations like max or averaging.

We use a CNN with two hidden convolutional layers. First, one takes the (3 x 24 x 50 ) price tensor as input and performs convolutions of size 1 x 3 to create a hidden layer consisting of 2 feature maps of size 24 x 48, which is followed by convolutions that lead to a final layer that feeds into the portfolio vector memory. 

\begin{figure}[H]
    \includegraphics[width=400pt]{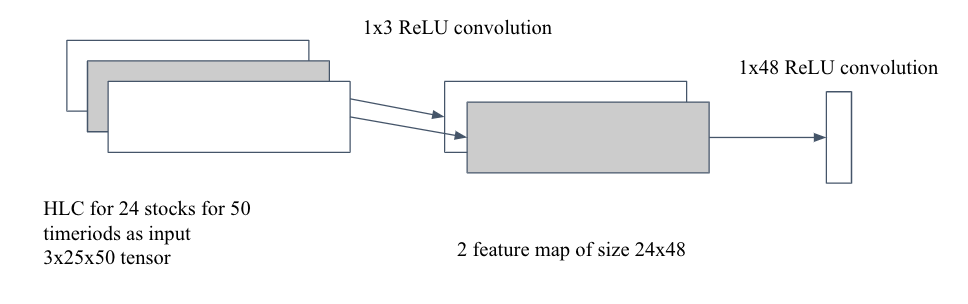}
    \caption{Schematic of the convolutional neural network (without the portfolio vector memory layer), the network takes the (3 x 24 x 50 ) price tensor as input and performs convolutions of size 1 x 3 to create a hidden layer consisting of 2 feature maps of size 24 x 48 which is followed by convolutions that lead to a final layer that feeds into the portfolio vector memory.}
    \label{fig:CNN}
\end{figure}

\begin{table}[H]
\centering
\begin{tabular}{|l|l|}
\hline
Network Type        & Convolutional Neural Network     \\\hline
Training Date Range & 1st Jan 2008 to 23rd March 2017  \\\hline
Testing Date Range  & 24rd March 2017 to 1st June 2020 \\\hline
Layers              & 1x2 Convolutional Layer          \\
                    & 1x24 Convolutional Layer         \\
                    & Portfolio Vector Memory Layer    \\\hline
Samples             & 4535                             \\\hline
Learning Rate       & 0.028                            \\\hline
Epochs              & 50000                            \\\hline
Batch Size          & 109                              \\\hline
Training Method     & Adam Optimizer      \\
\hline
\end{tabular}

\caption{Configuration of the CNN Network}
\end{table}

\subsubsection{Recurrent Neural Network}
Recurrent neural networks are variants of neural networks that are more suitable for sequential data. An RNN structure is different from a usual neural network as it has a recurrent connection or feedback with a time delay \cite{refRNN}. RNNs can use their memory to process variable-length sequences of inputs, making them applicable to tasks such as speech recognition and natural language processing. RNNs remember things learned from prior inputs while generating outputs. The output from an RNN is influenced not just by weights applied on inputs like a regular neural network, but also by a "hidden" state vector representing the context based on prior input/output, which makes the sequence of inputs critical to the output generated. 

We use a simple recurrent neural network with 20 units and 50 steps. We also add a dropout of 0.1 to avoid overfitting. The output from this network goes to the portfolio vector memory. 

\begin{figure}[H]
 
    \includegraphics[width=400pt]{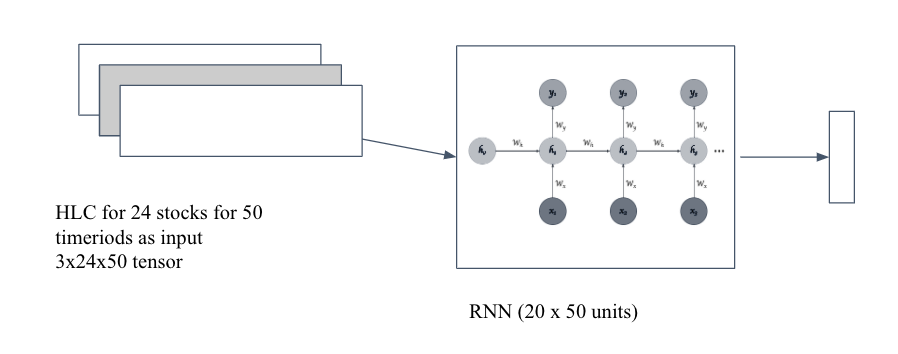}
    \caption{Schematic of the recurrent neural network architecture with 20 units and 50 steps which takes the input price tensor as input. The output from this network goes to the portfolio vector memory. }
    \label{fig:RNN}
\end{figure}

\begin{table}[H]
\centering
\begin{tabular}{|l|l|}
\hline
Network Type        & Recurrent Neural Network     \\\hline
Training Date Range & 1st Jan 2008 to 23rd March 2017  \\\hline
Testing Date Range  & 24rd March 2017 to 1st June 2020 \\\hline
Layers              & 20 units RNN Layer         \\
                    & Portfolio Vector Memory Layer    \\\hline
Samples             & 4535                             \\\hline
Learning Rate       & 0.00028                            \\\hline
Epochs              & 50000                            \\\hline
Batch Size          & 109                              \\\hline
Training Method     & Adam Optimizer      \\
\hline
\end{tabular}

\caption{Configuration of the RNN Network}
\end{table}

\subsubsection{Long Short Term Memory Neural Network}
A long short term memory network is a particular case of the recurrent neural network \cite{refLSTM}. This architecture also has a feedback loop and is useful for processing sequences that make it applicable in natural language processing and speech research. A standard LSTM unit comprises a cell, an input gate, an output gate, and a forget gate, which supports read, write and reset operations for the cells. The cell remembers values over arbitrary time intervals, and the three gates regulate the flow of information into and out of the cell. LSTMs were created to counter the vanishing and exploding gradient problems typical in RNNs, making it suitable for time series processing. There can be lags of unknown duration between important events in a time series.

We use a long short term memory network with 20 units and 50 steps. We also add a dropout of 0.1 to avoid overfitting. The output from this network goes to the portfolio vector memory. 
\begin{figure}[H]

    \includegraphics[width=400pt]{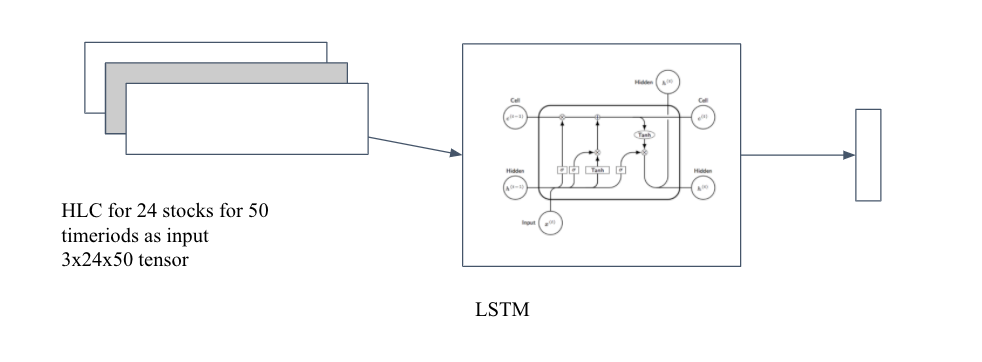}
    \caption{Schematic of the long short term memory neural network architecture with 20 units and 50 steps which takes the input price tensor as input. The output from this network goes to the portfolio vector memory.}
    \label{fig:LSTM}
\end{figure}

\begin{table}[H]
\centering
\begin{tabular}{|l|l|}
\hline
Network Type        & Long Short Term Memory     \\\hline
Training Date Range & 1st Jan 2008 to 23rd March 2017  \\\hline
Testing Date Range  & 24rd March 2017 to 1st June 2020 \\\hline
Layers              & 20 units LSTM Layer         \\
                    & Portfolio Vector Memory Layer    \\\hline
Samples             & 4535                             \\\hline
Learning Rate       & 0.0028                            \\\hline
Epochs              & 50000                            \\\hline
Batch Size          & 109                              \\\hline
Training Method     & Adam Optimizer      \\
\hline
\end{tabular}

\caption{Configuration of the LSTM Network}
\end{table}
\subsection{Transaction Costs and Portfolio Vector Memory}

Transactions in the financial market are not free. There are several costs related to trading \cite{costs}. These costs can be:
\begin{enumerate}
    \item Transaction cost that one pays to the broker, the exchange, the tax authority, and other intermediaries
    \item Market impact as the negative effect that a market participant has when it buys or sells an asset. 
\end{enumerate}

In this work, we are not considering market impact, assuming that the order sizes are too small to impact the market prices.
The transaction cost impacts the performance of the portfolio. In a free market, for a portfolio represented by asset weights $w^t$ (\textit{summing up to 1}) and asset prices $p^t$ and returns $r^t$, the portfolio $V^t$ value at time t is

\[ V^t = w^t.p^t \]

and at the next period is 

\[ V^{t+1} = V^t * [w^t . r^{t+1}]\]

In the presence of transaction costs, the value becomes

\[V^{t+1} = V^t * [[w^t . r^{t+1}] - \mu*|w^t - w^{t-1}_{rolled}|] \]

where $\mu$ is the transaction cost (assumed to be five basis points in our case)

If the weights $w^t$ are very volatile from one period to next, there would be a huge transaction cost component driving down returns. 
The portfolio vector memory component of our reinforcement learning network architecture addresses this. Inspired by \cite{jiang2017deep} and \cite{mnih2015humanlevel}, the final layer of our network consists of a portfolio vector memory, which includes the portfolio weights from the previous 20 periods along with current period prediction coming from the previous layers of the network.

\begin{figure}[H]
    \includegraphics[width=400pt]{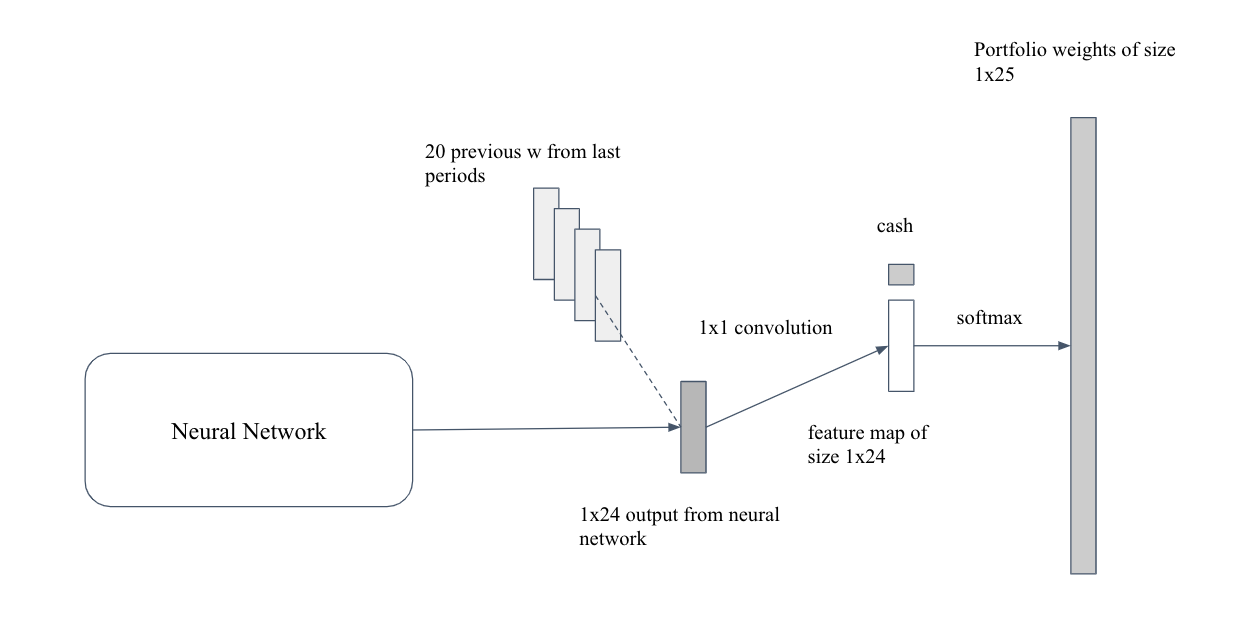}
    \caption{Schematic of the portfolio vector memory which includes the weights from previous 20 periods. The current network prediction is processed along with the previous weights with a single convolution to get the final feature of size 1x24 that goes through a softmax layer to get the final weights}
    \label{fig:PVM}
\end{figure}

This final layer in the network enables it to keep track of the portfolio turnover from one period to the next, incurring such costs only when necessary. The final layer is one of the most differentiating aspects of the architecture that makes it superior to the unsupervised and supervised learning frameworks, leading to good predictions but very noisy weights and high turnover.

\subsection{Reward Function}

Unlike a supervised learning approach where the reward function is generic, like the accuracy of prediction or mean square error, the reward function can be more holistic and entrenched in the reinforcement learning setup.

The reward that we have looked at maximizing is the average logarithmic cumulative return. The average algorithmic cumulative return can be represented as,

\[ R^T = 1/T * log(V^T/V^0) \]

which translates to 

\[ R^T = 1/T * \sum_{t=0}^{T} (r_t)\]

Where $R^T$ is the reward function at time $T$, $V^T$ is the value of the portfolio at time $T$, $r^t$ is the log return of the portfolio at time $t$.

This function considers each episodic return $r^t$ and is agnostic to the period for which it is run, as it averages over $T$. Each episode gets an equal weight, which makes this reward function all-inclusive - considering long-term and short-term returns.

\subsection{Learning} \label{topic:learning}

We define our \textit{policy} $ \pi $ as a mapping from state space to action space. 
\[ \pi : S \,\to\, A \]

The policy $J_T$ is defined by the reward function as mentioned above, is a function of network parameters $\theta$, action space $a_T=\pi_\theta(s_T)$

\[ J_T = R(s_1, \pi_\theta(s_1)....s_T, \pi_\theta(s_T))\]

We use gradient descent to update our parameters with the given learning rate ($\lambda$) in the direction of the gradient.

\[ \theta \,\to\, \theta + \lambda\bigtriangledown_\theta [J_{[t_0, t_f]}](\pi_\theta) \]

The gradient descent process allows us to reach the optimal parameters suited for the trading environment. 

\subsection{Comparative Frameworks} \label{traditional}

These are the frameworks we compare the reinforcement learning models to:
\begin{enumerate}
    \item \textbf{Equal Weighted Portfolio}
    
    Here the stocks have equal weights that are regularly rebalanced to be equal at each one-day interval. \cite{up}
    
   \item \textbf{Markowitz Approach - Mean Variance Optimization}
   
Here, the problem is cast as mean-variance optimization.
Mathematically, if there are $n$ assets with an expected return vector $\mu$ and covariance matrix $\Omega$, then the optimal
allocation $w$ is the solution to the following quadratic problem

\begin{equation}
{\displaystyle {{\arg_w \min }}\left(\frac{1}{2} \right) w^T \Omega w}
\end{equation}

subject to \[ \mu^T w >= \mu_b ,\]
and $\mu_b$ is the acceptable baseline expected rate of return.

The lookback for calculating expected returns is 50 periods (of one-day granularity), and the covariance matrix is also calculated based on the same period. We choose this lookback to keep the framework comparable to the RL framework.

\item \textbf{Risk Parity}

Now for this method known as risk parity, the optimization setup is as follows,

\begin{equation}
{\displaystyle {{\arg_w \min }}\sum _{i=1}^{n}\left[w_{i}-{\frac {\sigma (w)^{2}}{(\Omega w)_{i}N}}\right]^{2}}
\end{equation}

where \[ \sigma (w)=\sum _{i=1}^{n}\sigma _{i}(w)\]

Research \cite{riskparity} has shown that, in general, using the same input for both problems results in entirely
different portfolios. Using the mean-variance approach generally leads to concentrated positions in a small fraction of the assets. On the contrary, employing risk parity yields a very diversified portfolio \cite{risk_parity2}.

The lookback for calculating the covariance matrix is 50 periods (of one-day granularity) to keep the framework comparable to the RL framework.

\item \textbf{Minimum Variance}

This method minimizes the portfolio's volatility, and it would work best when all returns are the same, but there is variation in risk. We can lower risks without lowering returns. The optimization setup is as follows,

\begin{equation}
{\displaystyle {{\arg_w \min }} \left( w^T \Omega w\right)}
\end{equation}

The lookback for calculating the covariance matrix is chosen to be 50 periods (of one-day granularity). The lookback is chosen to keep the framework comparable to the RL framework.

\end{enumerate}

\section{Results}

\subsection{Traditional Models}
Looking at the results of the traditional models [\ref{traditional}] first, 
\begin{figure}[H]
\centering
    \includegraphics[scale=0.33]{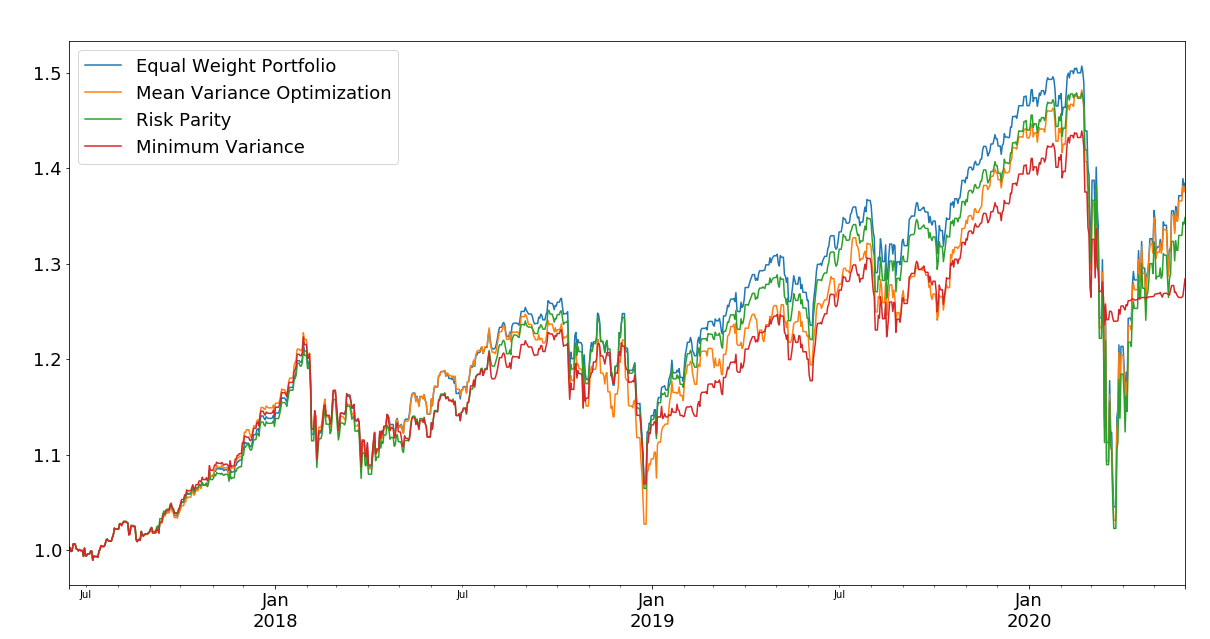}
    \caption{}
    \label{fig:returns_trad}
  \caption{Returns of the traditional strategies. The equal weight portfolio gives the best returns but the minimum variance portfolio gives the best Sharpe ratio and minimum drawdown. Results 24rd March 2017 to 1st June 2020. We apply 5 bps transaction costs}  
\end{figure}

\begin{table}[H]
\begin{adjustbox}{width=1\textwidth}
\begin{tabular}{|l|l|l|l|l|}
\hline
Algorithm & Total Returns                                            & Sharpe Ratio & Max Drawdown & Daily Turnover     \\
\hline
Equal Weight Portfolio     & 38.09 & 0.52 & 30.65 & 4.02  \\
Mean Variance Optimization & 37.74 & 0.51 & 30.43 & 2.98  \\
Risk Parity                & 34.51 & 0.49 & 30.88 & 2.00 \\
Minimum Variance                & 28.1 & 0.59 & 14.14 & 23.22 \\
\hline
\end{tabular}
\end{adjustbox}
\caption{Returns of the traditional strategies. The equal weighted portfolio gives the best returns but the minimum variance portfolio gives the best Sharpe ratio and minimum drawdown. Results 24rd March 2017 to 1st June 2020.}
\end{table}
Weights distribution of each of these models are as follows:
\begin{figure}[H]
\centering
    \includegraphics[scale=0.33]{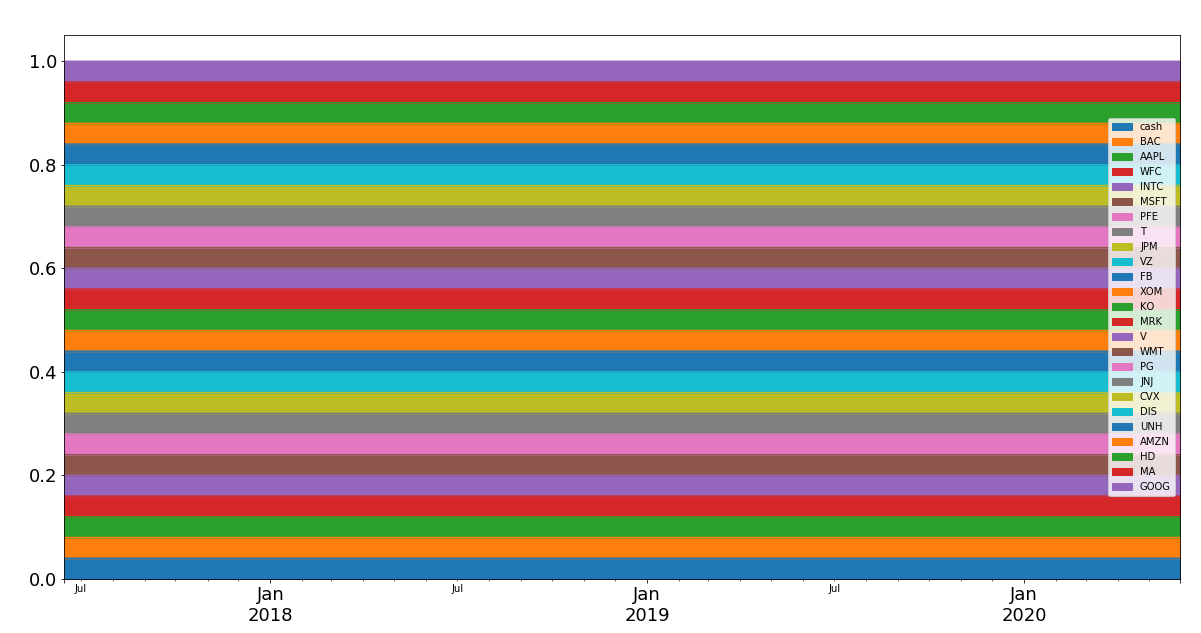}
    \caption{Weights for the equal weighted Strategy, as expected these are equal for all period. Results 24rd March 2017 to 1st June 2020.}
    \label{fig:eq_weight}
\end{figure}
\begin{figure}[H]
\centering
    \includegraphics[scale=0.33]{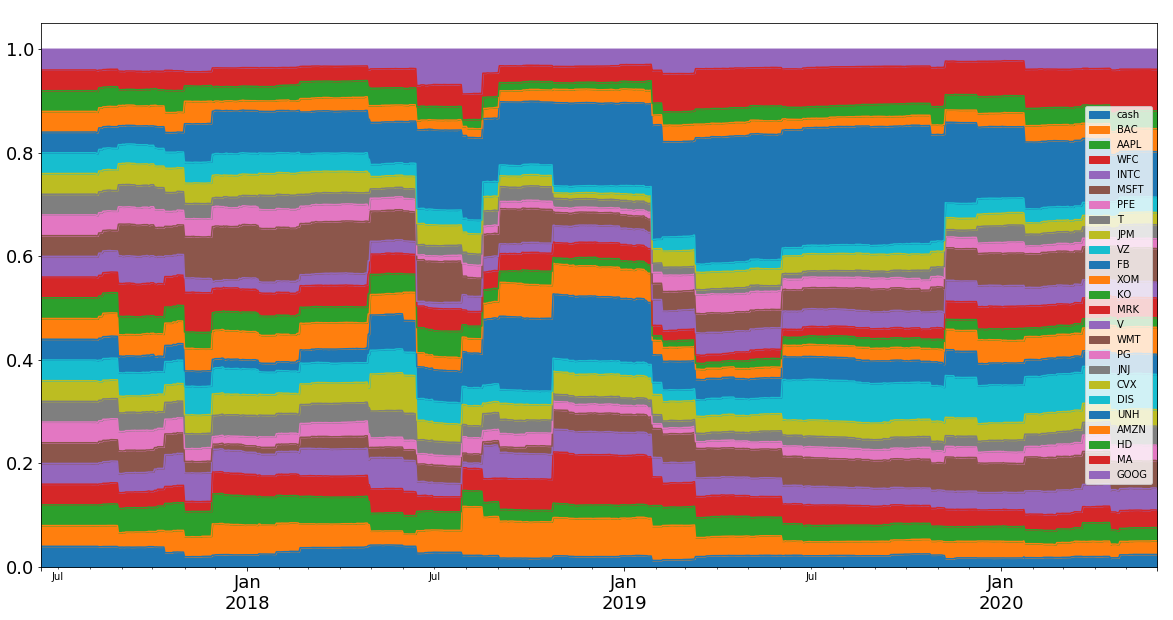}
    \caption{Weights for the mean variance optimal strategy, there is much more variation than the equal weight. }
    \label{fig:meanvar}
\end{figure}
\begin{figure}[H]
\centering
    \includegraphics[scale=0.33]{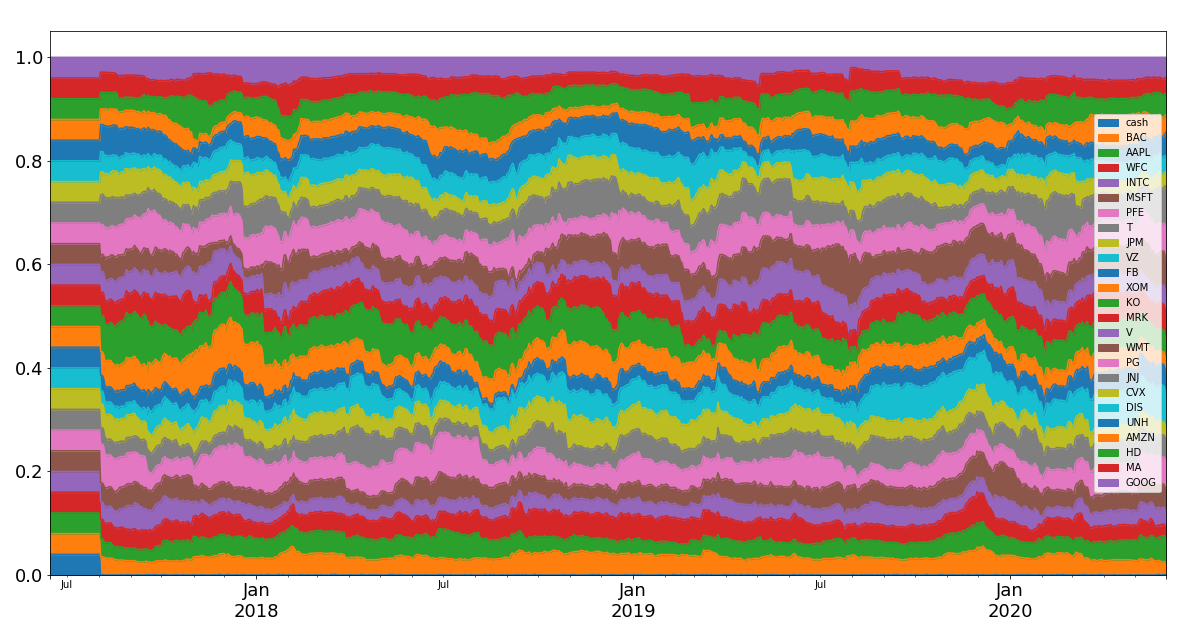}
    \caption{Weights for the risk parity strategy, these are closer to equal weight than mean variance optimal. Results 24rd March 2017 to 1st June 2020.}
    \label{fig:risk_parity}
\end{figure}
\begin{figure}[H]
\centering
    \includegraphics[scale=0.33]{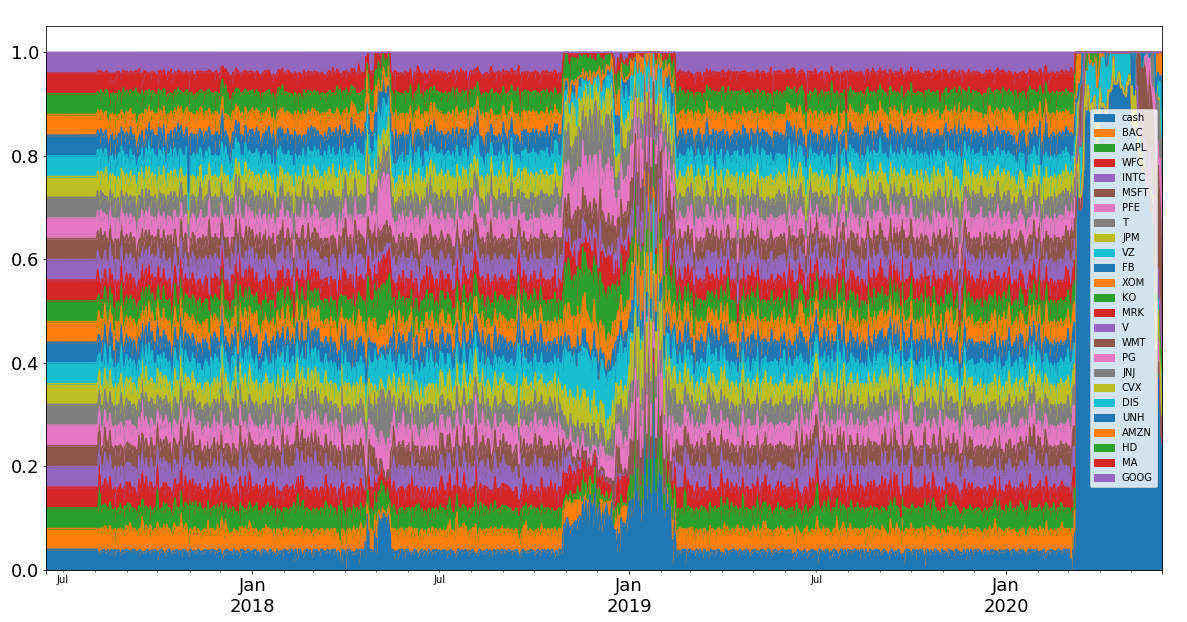}
    \caption{Weights for the minimum variance strategy, the weights deallocate to cash in extreme risk cases like March 2020. Results 24rd March 2017 to 1st June 2020.}
    \label{fig:mv_wts}
\end{figure}
\subsection{Convolutional Neural Network}
Looking at our first Neural Network the Convolutional Neural Network, we evaluate it with and without turnover control, we see that the returns without weight control are much higher than the traditional model and the turnover is quite high, using weight control using the PVM the returns are lower but the weights do not swing too wildly.
\begin{table}[H]
\begin{adjustbox}{width=1\textwidth}
\begin{tabular}{|l|l|l|l|l|}
\hline
Algorithm & Total Returns                                            & Sharpe Ratio & Max Drawdown & Daily Turnover     \\
\hline
CNN   &    39.56    & 0.52 & 31.79  & 6.69    \\
CNN No Weight Control  &    154.25    & 1.0 & 34.1  & 23.67    \\
\hline
\end{tabular}
\end{adjustbox}
\caption{Returns of the CNN model with and without turnover control with 5 bps cost. We see that the returns without turnover control are much higher than the traditional model but with weight control they are similar. Results 24rd March 2017 to 1st June 2020.}
\end{table}

\begin{figure}[H]
\centering
    \includegraphics[scale=0.33]{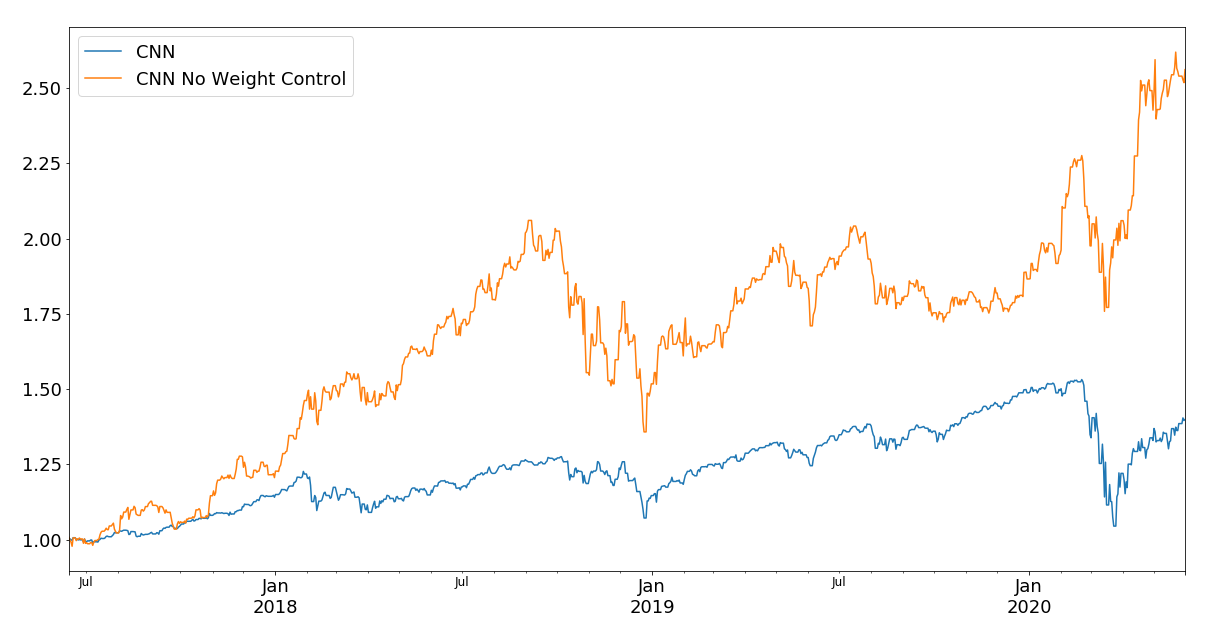}
    \caption{Returns of the CNN model with and without turnover control with 5 bps cost. With turnover control results are similar to traditional models, without turnover control they are much higher. Results 24rd March 2017 to 1st June 2020.}
    \label{fig:cnn_ret}
\end{figure}

\begin{figure}[H]
\centering
    \includegraphics[scale=0.33]{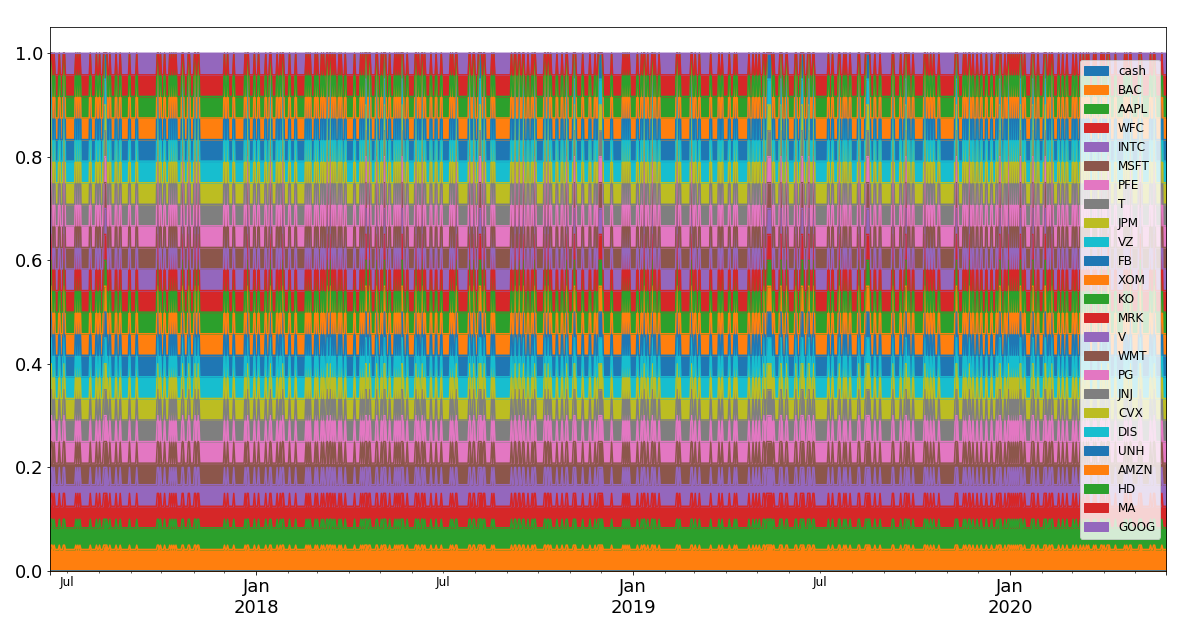}
    \caption{Weights for CNN with turnover control and cost of 5 bps, we see that the weights not varying wildly and are well diversified. Results 24rd March 2017 to 1st June 2020.}
    \label{fig:cnn_wt_noturnover}
\end{figure}
\begin{figure}[H]
\centering
     \includegraphics[scale=0.33]{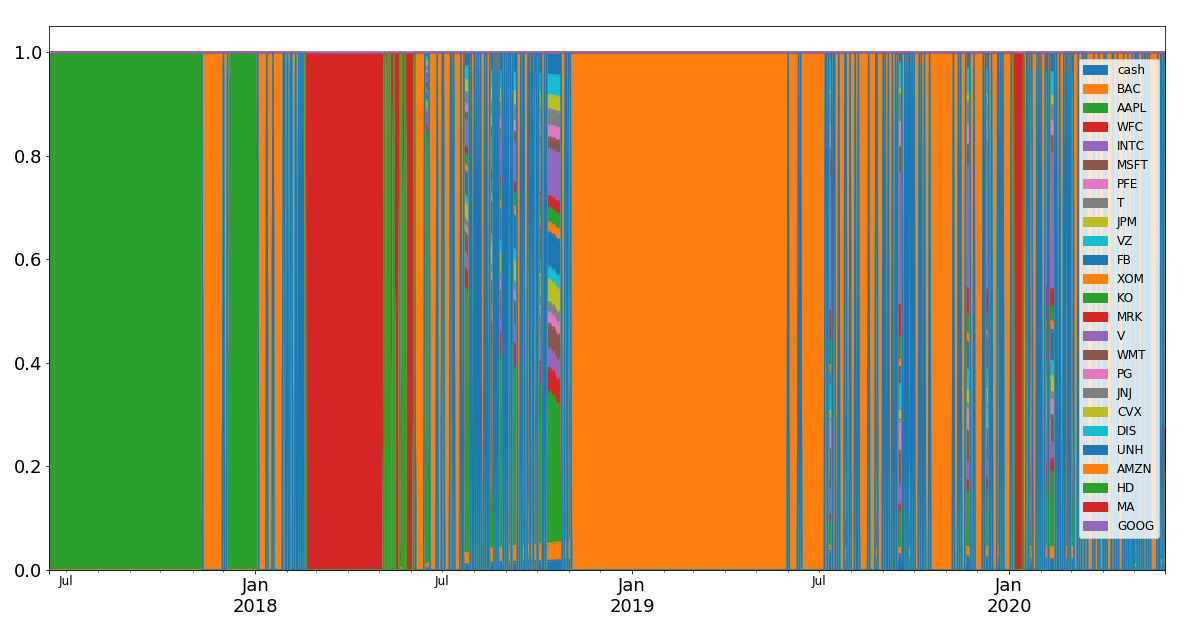}
    \caption{Weights for CNN without turnover control with Cost  of 5 bps, we see that the weights are very wildly varying to be fully allocated to the best performing stock.Results 24rd March 2017 to 1st June 2020.}
    \label{fig:cnn_wt}
\end{figure}

We can see that applying turnover control minimizes idiosyncratic \& stock-specific risk by not being fully allocated to a single stock. The control also minimizes turnover cost, which is preferable for long-only investors. We stick to using turnover control for the other networks.

\subsection{Recurrent Neural Network}

Now let us look at how the RNN model (with turnover control) does.

\begin{table}[H]
\begin{adjustbox}{width=1\textwidth}
\begin{tabular}{|l|l|l|l|l|}
\hline
Algorithm & Total Returns                                            & Sharpe Ratio & Max Drawdown & Daily Turnover     \\
\hline
RNN         & 53.92  & 0.53  & 30.87 & 23.16     \\
\hline
\end{tabular}
\end{adjustbox}
\caption{Returns of the RNN model with cost of 5 bps. The returns are lower than the CNN model but higher than the traditional models}
\end{table}
\begin{figure}[H]
\centering
    \includegraphics[scale=0.33]{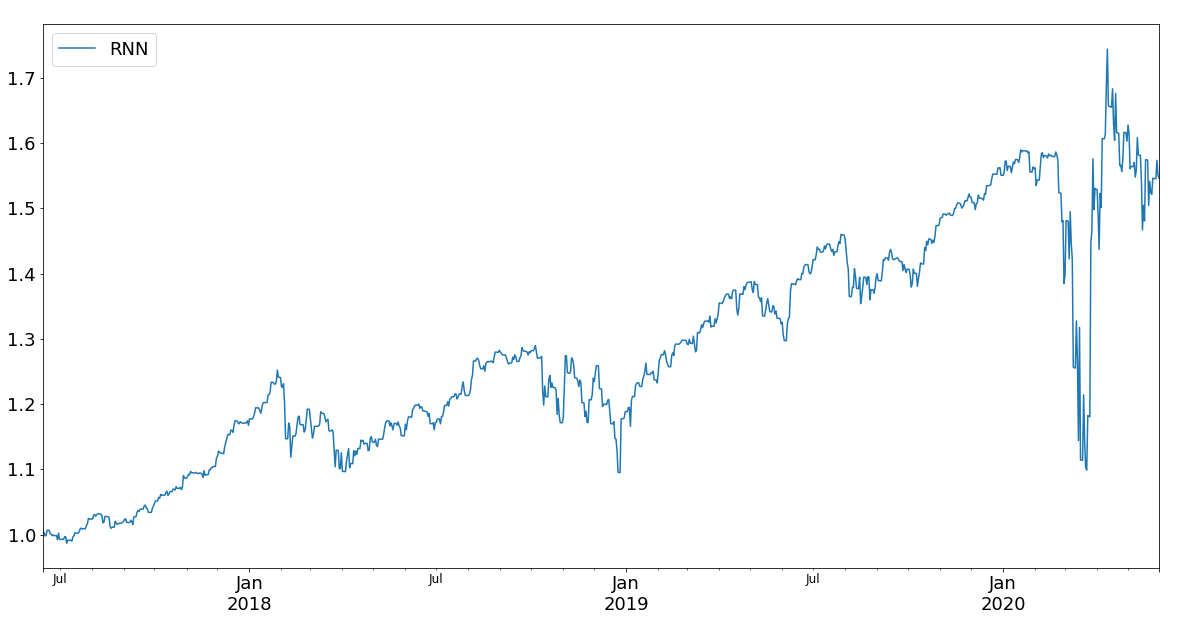}
    \caption{Returns for RNN model with cost of 5 bps. These are much lower than the CNN model but higher than traditional models}
    \label{fig:rnn_ret}
\end{figure}
\begin{figure}[H]
\centering
    \includegraphics[scale=0.33]{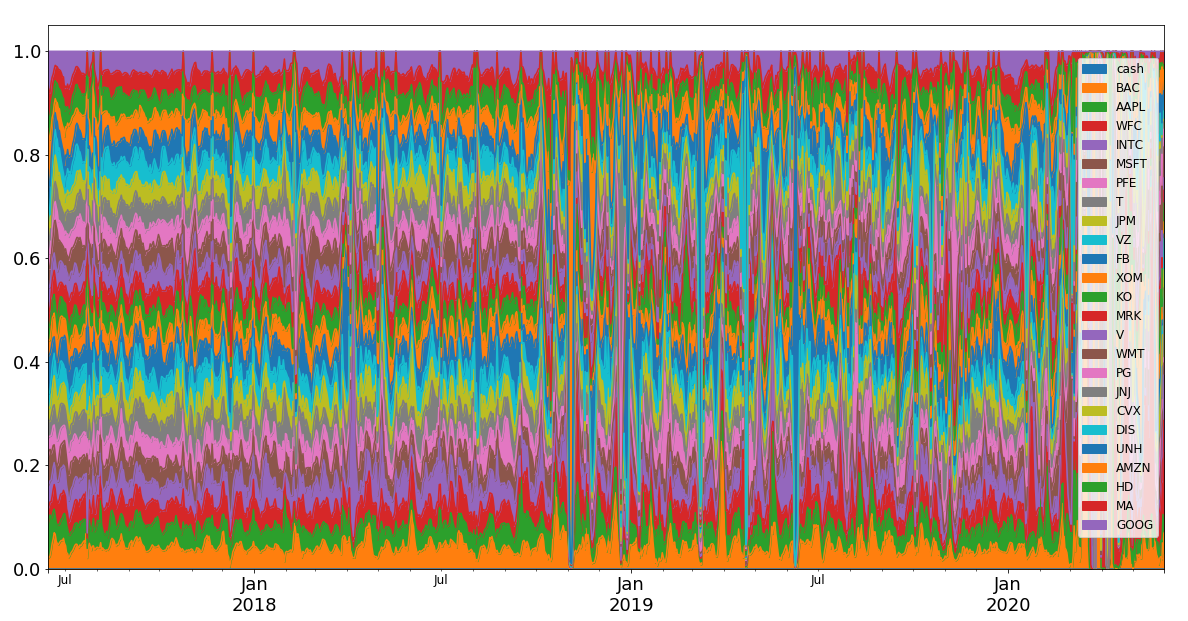}
    \caption{Weights for RNN model with cost of 5 bps. The weights vary more than traditional models and a wilder move is scene in risk conditions like March 2020. Results 24rd March 2017 to 1st June 2020.}
    \label{fig:rnn_wt}
\end{figure}
\subsection{Long Short Term Memory Neural Network}

Looking at the LSTM model (with turnover control) returns.
\begin{table}[H]
\begin{adjustbox}{width=1\textwidth}
\begin{tabular}{|l|l|l|l|l|}
\hline
Algorithm & Total Returns                                            & Sharpe Ratio & Max Drawdown & Daily Turnover     \\
\hline
LSTM         & 55.76  & 0.63  & 29.1 & 12.41    \\
\hline
\end{tabular}
\end{adjustbox}
\caption{Returns of the LSTM model with cost of 5 bps. The returns are higher than RNN but lower than CNN. The turnover is the lowest in this model among RL models}
\end{table}
\begin{figure}[H]
\centering
    \includegraphics[scale=0.33]{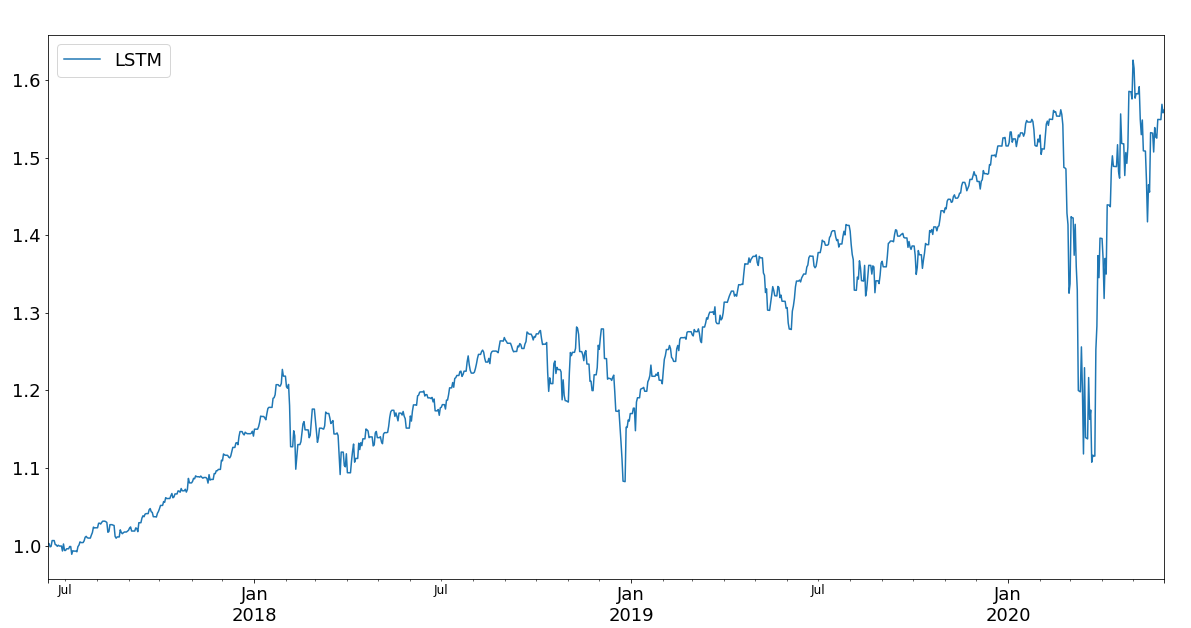}
    \caption{Returns for LSTM with Cost of 5 bps.  The returns are lower than CNN models but better than RNN}
    \label{fig:lstm_ret}
\end{figure}
\begin{figure}[H]
\centering
    \includegraphics[scale=0.33]{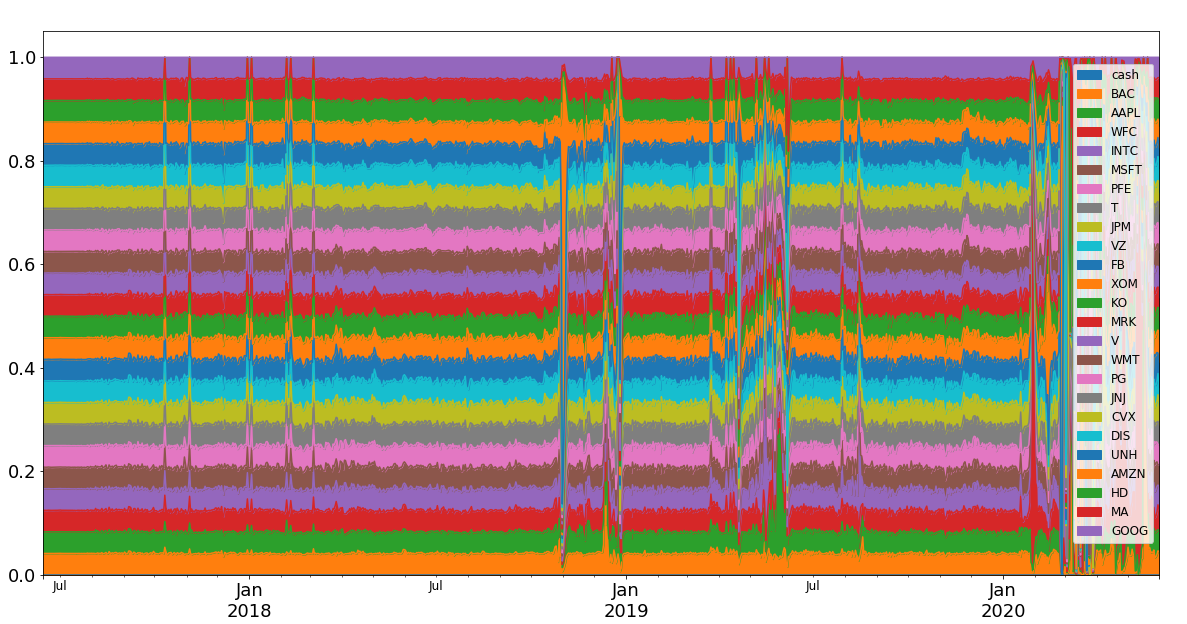}
    \caption{Weights for LSTM with Cost of 5 bps.  We see that the weights are close to equal weighted at most times with large variations in risk conditions like March 2020.Results 24rd March 2017 to 1st June 2020. }
    \label{fig:lstm_wt}
\end{figure}
\section{Main Results}

\begin{figure}[H]
\centering
    \includegraphics[scale=0.33]{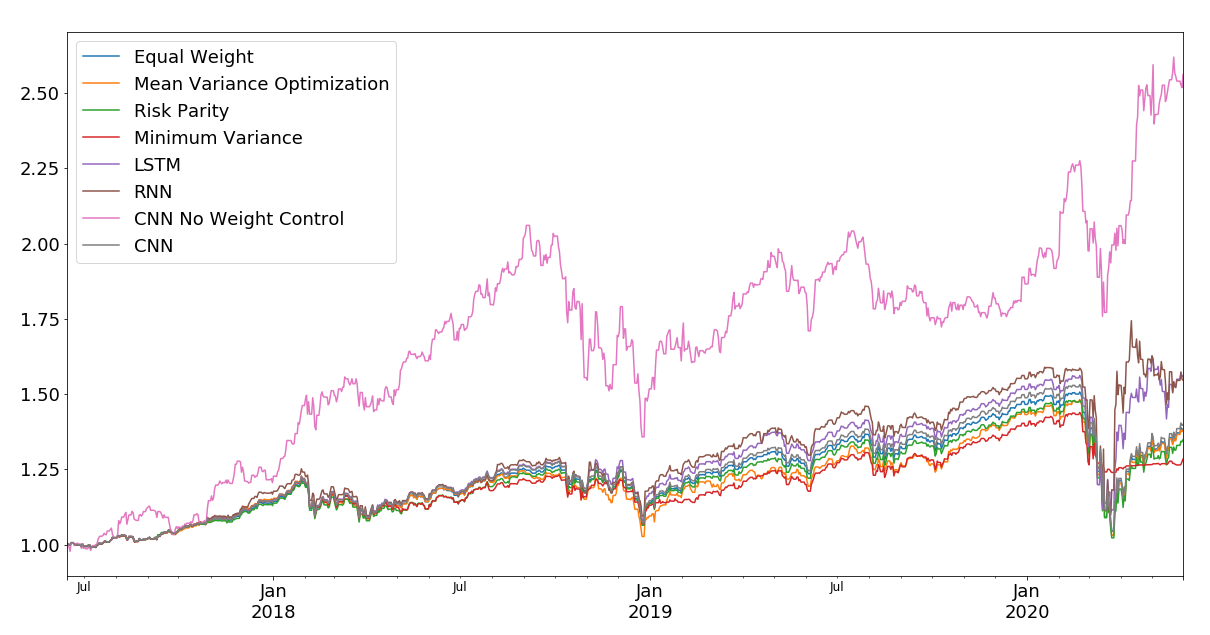}
    \caption{Returns for all the models together. CNN without turnover control here gives the best cost adjusted returns followed by LSTM with turnover control. All the neural networks perform better than the traditional models.Results 24rd March 2017 to 1st June 2020.}
    \label{fig:returns_all}
\end{figure}

\begin{table}[H]
\begin{adjustbox}{width=1\textwidth}
\begin{tabular}{|l|l|l|l|l|}
\hline
Algorithm & Total Returns                                            & Sharpe Ratio & Max Drawdown & Daily Turnover     \\
\hline
Equal Weight Portfolio     & 38.09 & 0.52 & 30.65 & 4.02  \\
Mean Variance Optimization & 37.74 & 0.51 & 30.43 & 2.98  \\
Risk Parity                & 34.51 & 0.49 & 30.88 & 2.00 \\
Minimum Variance                & 28.1 & 0.59 & 14.14 & 23.22 \\
CNN   &    39.56    & 0.52 & 31.79  & 6.69    \\
CNN No Weight Control  &    154.25    & 1.0 & 34.1  & 23.67    \\
RNN         & 53.92  & 0.53  & 30.87 & 23.16     \\
LSTM         & 55.76  & 0.63  & 29.1 & 12.41    \\
\hline
\end{tabular}
\end{adjustbox}
\caption{Returns for all the models together. CNN without turnover control here gives the best cost adjusted returns followed by LSTM. All the neural networks perform better than the traditional models. Also notable is that the turnover is not too high for the neural networks when compared to the traditional models. Results 24rd March 2017 to 1st June 2020.}

The reinforcement learning-based agents perform better than the traditional models in the period considered, and they also have a lower turnover.

Based on the results, this framework shows to be a valuable framework for the portfolio management problem. It learns from the data to maximize the returns 
considering risks and transaction costs.

\end{table}

\section{Conclusion}

We use a deep reinforcement learning framework to asset allocate US stocks using different deep learning architectures: Long Short Term Memory Networks, Convolutional Neural Networks, and Recurrent Neural Networks in an almost model-free framework, and we compare them with traditional asset allocation approaches like mean-variance, minimum variance, risk parity and equally weighted.  All these traditional asset allocation approaches require a prediction step and optimization step. Our approach deals with prediction and optimization, including transaction costs.
Our Deep Reinforcement Learning approach shows better results than traditional approaches for the top 24 US stocks using a simple reward function. In Finance, no training to test error generalization results are guaranteed. We can say, though, that the Deep Reinforcement Learning modeling framework can deal with time series prediction and portfolio allocation, including transaction costs. 
 It is remarkable the fact that Deep Reinforcement Learning is being given only the time series of the assets, and it learns to deal with returns and risk without risk being given a feature or considered in the reward function, showing again the flexibility of reinforcement and deep learning methods to deal with almost model-free tasks. In this problem, they also have to deal with the non-trivial behavior of financial time series.
We will research in the future with different reward functions, including risk on the reward function is a natural step, including market impact in addition to linear transaction costs, also if the framework can learn diversification and idiosyncratic risk. We will focus on the interpretability of the different reinforcement learning approaches compared to traditional approaches as well as using exogenous factors in the state space and the time series.

\bibliographystyle{apalike}
\bibliography{references}

\end{document}